\documentclass{pasj01}
\usepackage{csquotes}
\usepackage{graphicx}	
\usepackage{booktabs}
\usepackage{lineno}
\usepackage[caption=false]{subfig}
\Received{$\langle$2021 July 5$\rangle$}
\Accepted{$\langle$2021 November 14$\rangle$}
\Published{$\langle$publication date$\rangle$}

\begin{document}

\title{Theoretical study of infrared spectra of interstellar PAH molecules with N, NH \& NH$_2$ incorporation}
\author{Akant \textsc{Vats}\altaffilmark{1}%
\thanks{}}
\altaffiltext{1}{Department of Physics, Banaras Hindu University, Varanasi- 221005, India}
\email{ekantvv@gmail.com}

\author{Amit \textsc{Pathak}\altaffilmark{1}}
\email{amitpah@gmail.com}
\author{Takashi \textsc{Onaka}\altaffilmark{2,3}}
\email{onaka@astron.s.u-tokyo.ac.jp}
\author{Mridusmita \textsc{Buragohain}\altaffilmark{2}}
\author{Itsuki \textsc{Sakon}\altaffilmark{2}}
\author{Izumi \textsc{Endo}\altaffilmark{2}}

\altaffiltext{2}{Department of Astronomy, University of Tokyo, Tokyo 113-0033, Japan}
\altaffiltext{3}{Department of Physics, Faculty of Science and Engineering, Meisei University, 2-1-1, Hodokubo, Hino, Tokyo 191-8506, Japan}


\maketitle

\begin{abstract}
This work presents theoretical calculations of infrared spectra of nitrogen (N)-containing polycyclic aromatic hydrocarbon (PAH) molecules with incorporation of N, NH and NH$_2$ using density functional theory (DFT). The properties of their vibrational modes in 2--15 $\mu \rm m$ are investigated in relation to the Unidentified Infrared (UIR) bands. It is found that neutral PAHs, when incorporated with NH$_2$ and N (at inner positions), produce intense infrared bands at 6.2, 7.7 and 8.6 $\mu \rm m$ that have been normally attributed to ionized PAHs so far. The present results suggest that strong bands at 6.2 and 11.2 $\mu \rm m$ can arise from the same charge state of some N-containing PAHs, arguing that there might be some N-abundant astronomical regions where the 6.2 to 11.2 $\mu \rm m$ band ratio is not a direct indicator of PAHs' ionization. PAHs with NH$_2$ and N inside the carbon structure show the UIR band features characteristic to star-forming regions as well as reflection nebulae (Class A), whereas PAHs with N at the periphery have similar spectra to the UIR bands seen in planetary nebulae and post-AGB stars (Class B). The presence of N atom at the periphery of a PAH may attract H or H$^{+}$ to form N-H and N-H$_2$ bonds, exhibiting features near 2.9--3.0 $\mu \rm m$, which are not yet observationally detected. The absence of such features in the observations constrains the contribution of NH and NH$_2$ substituted PAHs that could be better tested with concentrated observations in this range. However, PAHs with N without H either at the periphery or inside the carbon structure do not have the abundance constraint due to the absence of 2.9--3.0 $\mu \rm m$ features and are relevant in terms of positions of the UIR bands. Extensive theoretical and experimental studies are required to obtain deeper insight.
\end{abstract}
\section{Introduction}
The presence of polycyclic aromatic molecules in the interstellar medium (ISM) is established by the ubiquitous observations of the unidentified infrared (UIR) emission bands at 3.3, 6.2, 7.7, 8.6, 11.2, 12.7 and 16.4 $\mu \rm m$ (3030, 1610, 1280, 1150, 885, 787 and 609 cm$^{-1}$) towards various Galactic and extra-galactic sources (Gillet et al. 1973; Cohen et al. 1986; Li 2020). These bands have been ascribed to infrared fluorescence of polycyclic aromatic hydrocarbon (PAH) molecules excited by UV and optical photons (Leger \& Puget 1984; Allamandola et al. 1985). The UIR bands exhibit diversity in terms of peak position, width and intensity depending on the local environment of the ISM (Hony et al. 2001; Peeters et al. 2003; Sakon et al. 2004; Tielens 2008). These variations suggest the presence of varying forms of PAHs. Neutral PAHs have been proposed to produce strong 3.3 and 11.2 $\mu \rm m$ bands whereas the emissions at 6.2, 7.7 and 8.6 $\mu \rm m$ have been attributed to ionized PAHs (Schutte et al. 1993; Peeters et al. 2002; Tielens 2008; Schmidt et al. 2009). Besides this, several complex organics having disorganized structures show the potential to emit the UIR bands. These include Hydrogenated Amorphous Carbon (HAC) (Jones et al. 1990; Jones et al. 2017), Quenched Carbonaceous Composites (QCCs) (Sakata et al. 1987), coal (Guillois et al. 1996; Papoular et al. 1989)  and Mixed Aromatic/Aliphatic Organic Nanoparticles (MAONs) (Kwok \& Zhang 2011; Kwok \& Zhang 2013).  

The 5--9 $\mu \rm m$ region shows major variations from source to source with bands at 5.2, 5.7, 6.0, 6.2, 6.8, 7.7 and 8.6 $\mu \rm m$, in which the 7.7 $\mu \rm m$ band is stronger than others (Tielens et al. 2008). Peeters et al. (2002) found that the 7.7 $\mu \rm m$ band consists of subfeatures at $\sim$7.6 and $\sim$7.8 $\mu \rm m$, while the peak position of the 6.2 $\mu \rm m$ band lies either at $\sim$6.2 $\mu \rm m$ or at a slightly redder position of $\sim$6.3 $\mu \rm m$. The varying size of PAHs can successfully explain the emission at 6.3 $\mu \rm m$, while polycyclic aromatic nitrogen heterocycle (PANH) cations have been attributed to reproduce the 6.2 $\mu \rm m$ feature (Hudgins et al. 2005). Importance of PANHs has been noticed in photon-dominated regions (PDRs) as spectra of their cations are required to fit the 6.2 and 11.0 $\mu \rm m$ observed features towards NGC 7023 (Boersma et al. 2013). Nitrogen has large electronegativity that affects the dipole derivatives of C-C stretching modes and changes the IR characteristic of these modes. This induces the shifting of the 6.2 $\mu \rm m$ band towards higher frequency, which becomes prominent when N is incorporated deeper in PAHs (Hudgins et al. 2005; Tielens 2008).

In the ISM, the detection of benzo-nitrile (McGuire et al. 2018) and cyano-naphthalene (McGuire et al. 2021) strengthens the idea of existence of PANHs in the ISM. PAH and PANH cations (same structure and geometry) are found to be equally efficient in reaction with atomic H, and association of H at the nitrogen is more exothermic than association at the carbon (Demarais et al. 2014). If the lone-pair electrons of the nitrogen are not delocalized in the $\pi$ system, N atom has chance to make bond with ionized atomic H (Alvaro Galu\'e et al. 2010; Hudgins et al. 2005), leading to the formation of N-H and N-H$_{2}$. The high proton affinity of PANHs may also help in the formation of their protonated forms. The electronic spectra of protonated PANHs have also been studied experimentally for comparison with the diffuse interstellar bands (DIBs) (Noble et al. 2015).
\begin{figure*}
    \centering
     \includegraphics[width=.17\textwidth]{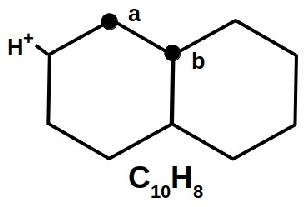}\hfill
   \includegraphics[width=.13\textwidth]{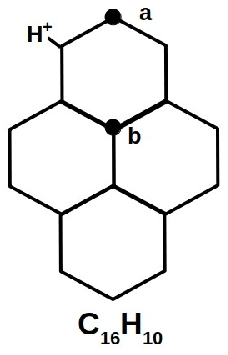}\hfill
   \includegraphics[width=.16\textwidth]{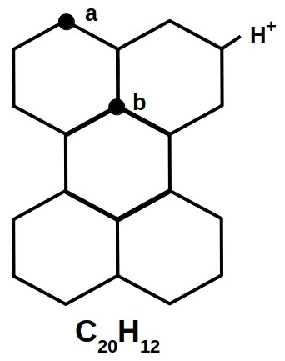}\hfill
    \includegraphics[width=.18\textwidth]{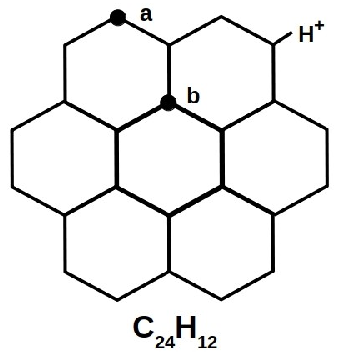}\hfill\\
     \includegraphics[width=.32\textwidth]{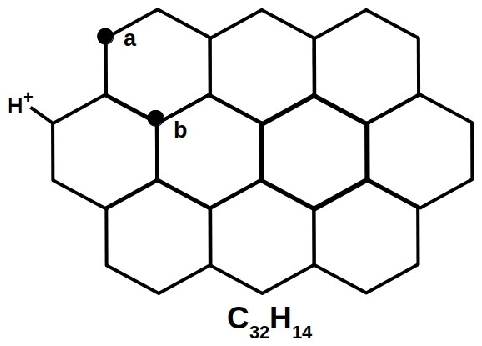}\hspace{15mm}
     \includegraphics[width=.39\textwidth]{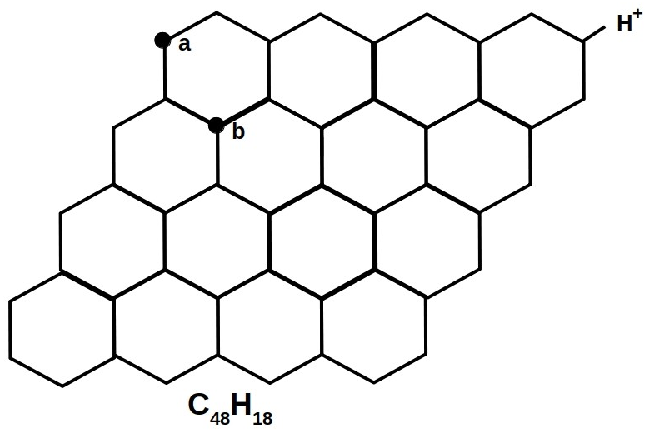}\\
     \includegraphics[width=.335\textwidth]{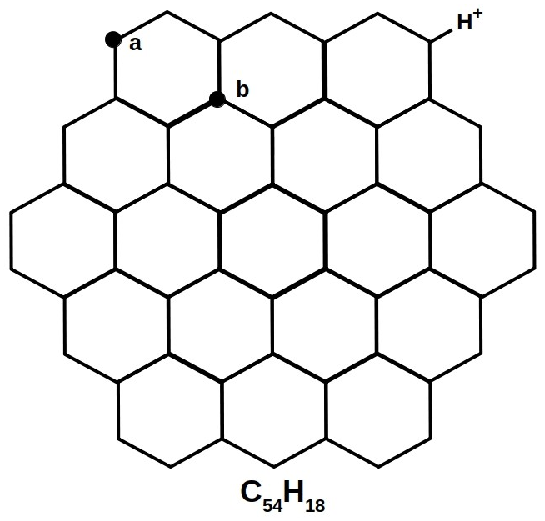}
    \caption{PAHs with N-substituted variants studied in this work. The black dots show the site of nitrogen (\enquote*{a} and \enquote*{b}); for \enquote*{a}, CH is replaced by N at the periphery (exo N-PAH) and for site \enquote*{b}, N substitutes a C inside the structure (endo N-PAH). Two other possibilities are considered for H bonding with the peripheral N (site \enquote*{a}) --- single H atom bonded with N (NH-PAH) and two H atoms bonded with N (NH$_2$-PAH). H$^+$ shows the protonation site.}
    \label{fig:my_label}
\end{figure*}

IR spectra of PANHs have been reported in context to the UIR bands, experimentally for small neutrals and cations by Mattioda et al. (2003) and for protonated forms (with H$^{+}$ at N) by Alvaro Galu\'e et al. (2010) and theoretically for large cations by Hudgins et al. (2005). Theoretical IR spectra of PAHs with CN, NH and NH$_2$ as side group have been studied for naphthalene (C$_{10}$H$_{8}$) and anthracene (C$_{14}$H$_{10}$) (Bauschlicher et al. 1998). The study concludes that PAHs with N-containing side groups may not be abundant and will not contribute significantly to the UIR bands based on the absence of N-H and C-N stretching features in astronomical spectra, while N in the PAH ring structure as proposed by Hudgins et al. (2005); Alvaro Galu\'e et al. (2010) and Noble et al. (2015) remains as potentially promising carriers of the UIR bands.

Calculations have been done for four variants of these PAHs with N, NH and NH$_2$ substitutions in neutral, cation and protonated forms to present a systematic study of the effects of nitrogen incorporation in a large samples of PAHs in this paper. This work includes medium and large sized PAHs (having up to 54 carbon atoms) that are relevant to be present in the interstellar space. Species with NH and NH$_2$ within the PAH ring along with their protonated (with H$^{+}$ at C) forms are reported for the first time. Section 2 describes the structure and the theoretical methodolgy. The results are presented and discussed in Section 3 and the astrophysical implications are given in Section 4.
\section{Calculation Method}
Density functional theory (DFT) is an appropriate computational approach to simulate the quantum states of PAH molecules, which is used extensively to investigate the IR spectra in the astrophysical context, e.g., Langhoff (1996); Hudgins et al. (2004); Pathak and Rastogi (2005); Pathak and Rastogi (2006); Pathak and Rastogi (2007); Pauzat et al. (2011); Buragohain et al. (2020). We employ the DFT calculations using Gaussian 09 with the B3LYP/6-31G++(d, p) basis set to optimize the geometries of PAHs that are further used to calculate the harmonic frequencies and IR intensities. Mode-dependent scaling factors have been used to scale the theoretical frequencies in order to bring them in accordance with experiments. A single scaling factor may not be sufficient as different vibrational motions require different values of the scaling factors. Here, the scaling factors are taken from Buragohain et al. (2015), which were determined after comparing the calculations for selected PAHs with experimental data. These are 0.974 for C-H out of plane bending modes, 0.972 for C-H in plane and C-C stretching modes and 0.965 for C-H stretching modes. 
\begin{figure*}
    \centering
   \subfloat{\includegraphics[height=22.2cm, width=18.2cm]{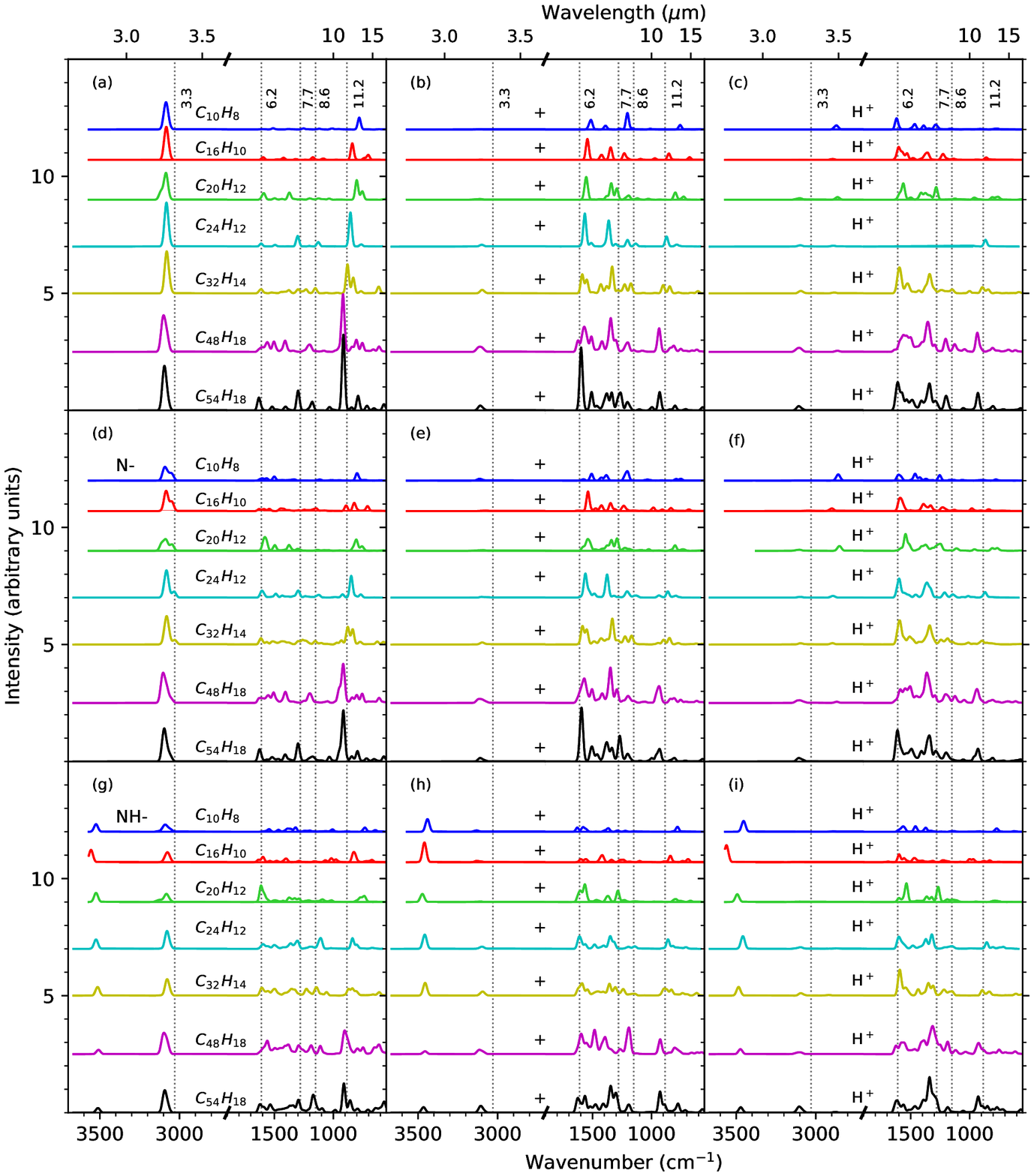}}
   \caption{Theoretical IR spectra of 0-15 $\mu \rm m$ region for naphthalene (blue), pyrene (red), perylene (green), coronene (cyan), ovalene (yellow), C$_{48}$H$_{18}$ (magenta) and circumcoronene (black) in five variants---PAH (I row), exo N-PAH (II row), NH-PAH (III row), NH$_2$-PAH (IV row) and endo N-PAH (V row) in neutral (I column), cationic (II column) and protonated (III column) forms, where N denotes the nitrogen atom. Dotted vertical lines show the observed positions of the PAH bands. Spectra are plotted assuming the Gaussian profile with the FWHM of 30 cm$^{-1}$. The diagonal lines in the horizontal axes denote the removed section between 2700-1900 cm$^{-1}$ due to the absence of features in this region.}
\end{figure*}

\begin{figure*}
\ContinuedFloat
\subfloat{\includegraphics[height=14.7cm, width=18.2cm]{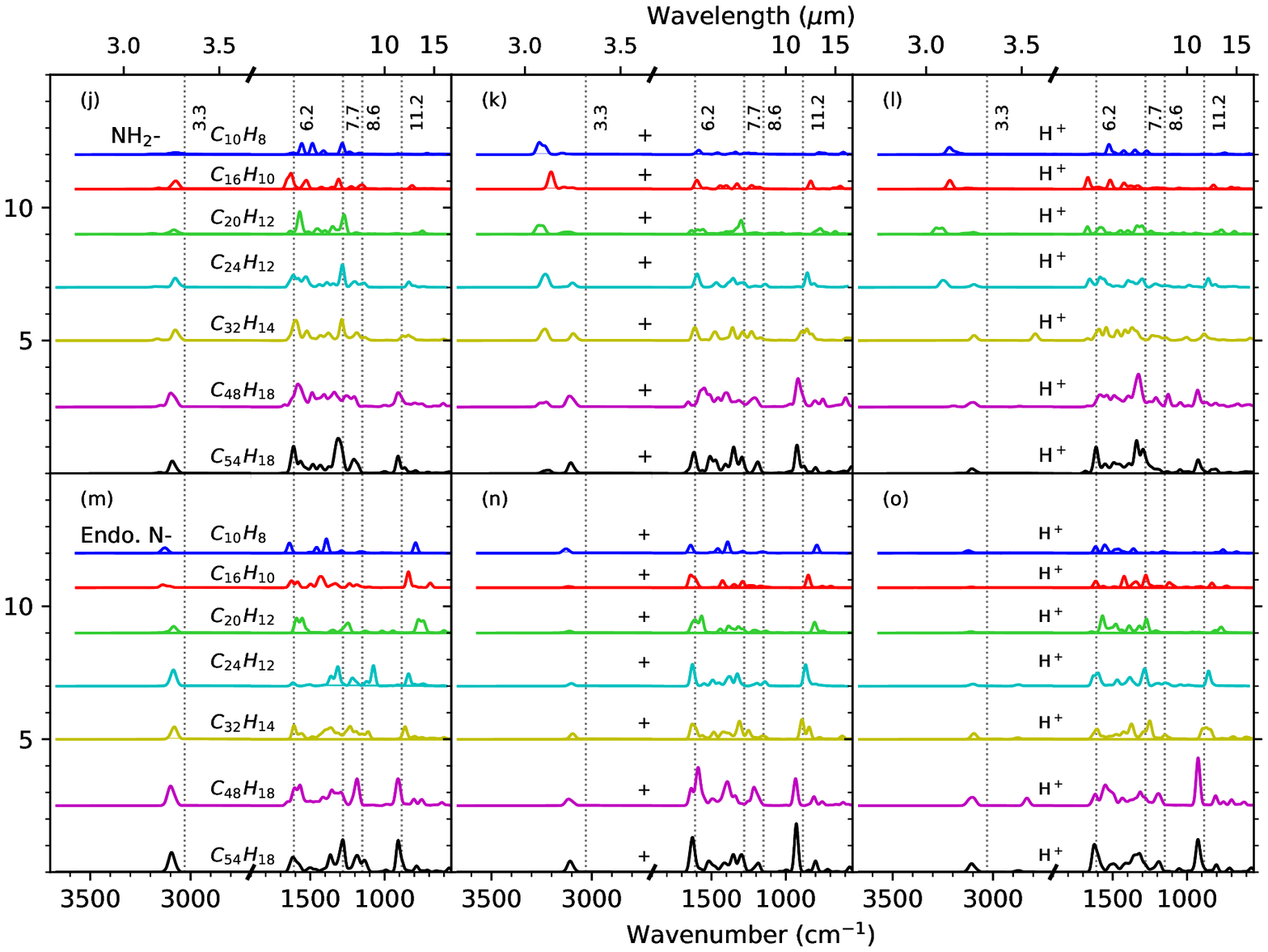}}
\caption{continued}
\end{figure*}

In this paper, we focus only on the position and intensity of the bands that correlate with the observed UIR bands. We are not taking into account the anharmonicity that affects the band profiles. For this purpose, DFT-B3LYP provides relevant results. Higher level theoretical methods like MP2, etc. are computationally very expensive especially for the case of large molecules such as those considered in this paper. Considering the fact that the accuracy of the results at B3LYP level for the band wavelengths (when corrected by the scaling factors) and the band strengths is not compromised in comparison to the computational resources required for MP2 calculations, therefore, the present results should be useful to signify a general outlook of N-containing PAHs. 

Seven PAHs have been considered (Figure 1), namely naphthalene (C$_{10}$H$_{8}$), pyrene (C$_{16}$H$_{10}$), perylene (C$_{20}$H$_{12}$), coronene (C$_{24}$H$_{12}$), ovalene (C$_{32}$H$_{14}$), C$_{48}$H$_{18}$ and circumcorornene (C$_{54}$H$_{18}$) with substitution of N, NH and NH$_2$ in four different N-incorporated variants. At site \enquote*{a} in Figure 1, N replaces CH with the lone pair on N atom not being part of the $\pi$ system (exoskeletal or exo N-PAH). This peripheral N atom may be bonded with one or two H atoms giving rise to NH-PAH and NH$_2$-PAH respectively. N may also be at site \enquote*{b} incorporated within the PAH structure (endoskeletal or endo N-PAH). 

An emission model is employed to transform the theoretically computed absorption spectra of PAHs into emission spectra for a justifiable comparison with the observed UIR bands. The model is based on studies by Schutte et al. (1993), Cook and Saykally (1998), Pech et al. (2002) and Pathak and Rastogi (2008). The model considers a PAH in an interstellar radiation field, corresponding to blackbody temperature T = 40,000K. The PAH molecule becomes internally excited equivalent to a peak temperature (T$_{p}$) depending on the heat capacity of the PAH. The excited PAH then relaxes by emitting a cascade of photons associated with the vibrational modes of the PAH. The emitted energy is computed for $\Delta$T = 1K and is integrated over the range T$_{p}$ to 50K. Below 50K, the emitted energy is found to be insignificant. The emission model considers the rate of absorption of photons to calculate the emitted energy, and to produce the emission spectrum. This emitted energy is summed up over the whole distribution of photon absorption. The intensity unit for this model is $10^{-13}$W$C^{-1}${$\mu \rm m$}$^{-1}$ (Pech et al. 2002). The spectra are convolved with 30 cm$^{-1}$ FWHM as the chosen value is typical for the PAH emission (Allamandola et al. 1989). 

We have also studied the ionization potential (IP) and H loss (from C atom) energy for each studied variant. Ionization potential for PAHs ranges from 6 to 8 eV, which remains similar when N replaces CH in exo N-PAH. For NH-PAH and endo N-PAH, the range for IP is 4 to 5 eV. The variant where N is attached with two H atoms (NH$_2$-PAH) has the IP values from 5 to 6 eV. The value decreases with increasing size as C$_{10}$H$_{8}$ and its N-substituted variants show larger values (up to 8 eV) than C$_{54}$H$_{18}$ and its N-substituted counterpart (up to 6 eV) following the typical behavior of PAHs. The H loss energy is 4 to 5 eV for every variant. 
\begin{table*}
\tbl{Summary of the theoretical mid IR behavior of PANH variants along with PAHs in neutral, cationic($^+$) and protonated (H$^+$) forms.}
{\begin{tabular}{lcccc}
\noalign{\vskip3pt}
\hline
Species & \multicolumn{4}{c}{Normalized intensity (Int$\rm_{rel}$)}\\\cmidrule{2-5}
& NH stretching & C-H stretching & C-C stretching/C-H in-plane & CH oop$^\Upsilon$\\
& (2.8-3.10 $\mu \rm m$)$^\Pi$ & (3.21-3.23 $\mu \rm m$)$^\Pi$ & (6.11-8.89 $\mu \rm m$)$^\Pi$ & (10.76-13.94 $\mu \rm m$)$^\Pi$\\
\hline
PAHs & & very strong & weak & strong$^\P$\\
$^+$ & & weak & very strong & weak$^\Sigma$\\
H$^+$ & & weak & very strong & weak$^\Sigma$\\\hline
N-PAHs (exo) & & very strong & weak & strong$^\P$\\
$^+$ & & weak & very strong & weak$^\Sigma$\\
H$^+$ & & weak & very strong & weak$^\Sigma$\\\hline
NH-PAHs & weak & very strong & weak & strong$^\P$\\
$^+$ & moderate & weak & very strong & weak$^\Sigma$\\
H$^+$ & moderate & weak & very strong & weak$^\Sigma$\\\hline
NH$_2$-PAHs & weak & moderate & very strong & moderate\\
$^+$ & moderate & weak & strong & very strong\\
H$^+$ & moderate & weak & very strong & strong\\\hline
N-PAHs (endo) & & strong & strong & very strong\\
$^+$ & & weak & strong & very strong\\
H$^+$ & & weak & strong & very strong\\\hline
\hline
\end{tabular}}
\begin{tabnote}
The data is presented for naphthalene (C$_{10}$H$_8$), pyrene (C$_{16}$H$_{10}$), perylene (C$_{20}$H$_{12}$), coronene (C$_{24}$H$_{12}$), ovalene (C$_{32}$H$_{14}$), C$_{48}$H$_{18}$ and circumcoronene (C$_{54}$H$_{18}$).

$^\Pi$this paper.

very strong (0.8 $\leq$ Int$_{rel}$ $\leq$ 1.0), strong (0.4 $\leq$ Int$_{rel}$ $\leq$ 0.7), moderate (0.1 $\leq$ Int$_{rel}$ $\leq$ 0.4) and weak (Int$_{rel}$ $\leq$ 0.1) are defined according to the relative intensity (Int$_{rel}$), obtained by taking the ratio of all intensities to the maximum. 

$^\P$very strong for C$_{48}$H$_{18}$ and circumcoronene.

$^\Sigma$moderate for C$_{48}$H$_{18}$ and circumcoronene.

$^\Upsilon$C-H oop bending features follow their emergence on the account of CH-groups present in any PANH. oop stands for out-of-plane.
\end{tabnote} 
\end{table*}

In the present work, the site for nitrogenation and protonation has been chosen by calculating the ground state energy of structures having different unique positions of substitution in a PAH, from which the structure with the lowest energy is used for further analysis. Table 1 shows normalized intensities to summarize the mid IR behavior of the chosen species, whereas Table 2 and 3 are presented on an absolute intensity scale to study the size effect of PAHs and their N-containing variants.
\section{Results and Discussion}~
Figure 2 shows spectra of seven PAHs in four N-substituted variants (Figure 1) for neutral, cations and protonated forms calculated in the present study for naphthalenes (blue), pyrenes (red), perylenes (green), coronenes (cyan), ovalenes (yellow), C$_{48}$H$_{18}$ (magenta) and circumcoronenes (black). Each row represents different variants of chosen molecules---pure PAH (Figure 2, row I), exo N-PAH (Figure 2, row II), NH-PAH (Figure 2, row III), NH$_2$-PAH (Figure 2, row IV) and endo N-PAH (Figure 2, row V), respectively. Likewise, from left to right; columns depict the neutral, cationic and protonated form.
\subsection{The C-H Stretching Vibrations (3.2--3.3 $\mu \rm m$)}~
The IR behavior of C-H stretching features for every PANH variant is summarized in Table 1 and the spectra are shown in Figure 2. The C-H stretching region is very similar for all the PANH variants (Figure 2). This resemblance involves neutral, cationic and protonated molecular forms here in terms of the peak position and intensity, and no significant effect of nitrogenation is recognized in this region for any variant (Table 1). The C-H stretching feature shows a small blue shift ($\sim$0.02) upon ionization\footnote{Ions are used for referring to both the cations and protonated PAHs.}, which is true for every PANH variant. In the present sample, the intensity of the 3.3 $\mu \rm m$ feature increases with the size because of the increase of C-H bonds (Bauschlicher et al. 2008). This is followed by each PANH variant as well. 
\subsection{The C-C stretching/C-H in-plane bending Vibrations (6.0--9.0 $\mu \rm m$)}~
The IR behavior for each PANH variant is summarized in Table 1, while the peak positions and emitted intensities of the C-C stretching/C-H in-plane bending modes for each species studied here, are listed in Table 2 with C/N ratio, where the molecules are arranged in the order of increasing size. For the 6.2 $\mu \rm m$ band, the dominant band peaks in between 6.0-6.6 $\mu \rm m$, while for the 7.7 and 8.6 $\mu \rm m$ bands, the dominant features peak at 7.21-8.28 and 8.4-8.89 $\mu \rm m$ respectively. The C-C stretching/C-H in-plane bending region of exo N-PAH (Figure 2, row II) and NH-PAH (Figure 2, row III) variants exhibits the same behavior as that of pure PAHs (Figure 2, row I), where ions display stronger intensities compared to neutrals (Table 1). However, for NH$_2$-PAH (Figure 2, row IV), this region shows larger intensities for neutrals compared to ions (Figure 2 \& Table 1). The endo N-PAH variant (Figure 2, row V) behaves differently for this region---as the size increases, the intensity of the 6.2 $\mu \rm m$ band is higher for ions, whereas neutrals have the strong 7.7 and 8.6 $\mu \rm m$ bands (Table 2). The intensity of the 6.2 and 7.7 $\mu \rm m$ bands increases with size for the pure PAH and N-PAH (Figure 2).
\begin{center}
\begin{longtable}[hbt!]{lclcc}
\caption{Calculated peak positions ($\mu \rm m$) and intensities ($10^{-13}$W$C^{-1}${$\mu \rm m$}$^{-1}$) of the dominant 6.2, 7.7 and 8.6 $\mu \rm m$ bands for PANH variants in neutral, cationic ($^+$) and protonated (H$^+$) forms}\\
\noalign{\vskip3pt}
\hline
Molecules & C/N ratio & 6.2 (I) & 7.7 (I) & 8.6 (I)\\
\hline
\endfirsthead

\multicolumn{5}{c}%
{{\bfseries \tablename\ \thetable{} -- continued from previous page}} \\
\hline
Molecules & C/N ratio & 6.2 (I) & 7.7 (I) & 8.6 (I)\\
\hline
\endhead

\hline \multicolumn{5}{|r|}{{Continued on next page}} \\ \hline
\endfoot
\endlastfoot

Naphthalene &  &  &  & \\
N-C$_{10}$H$_{8}$ (Exo) (Figure 2-d-blue) & 9 & 6.22 (0.04) & 7.55 (0.02) & 8.81 (0.02) \\
NH-C$_{10}$H$_{8}$ (Figure 2-g-blue) & 9 & 6.45 (0.06) & 7.56 (0.13) & 8.43 (0.05) \\
NH$_2$-C$_{10}$H$_{8}$ (Figure 2-j-blue) & 9 & 6.43 (0.65) & 7.79 (0.73) & 8.42 (0.21) \\
N-C$_{10}$H$_{8}$ (Endo) (Figure 2-m-blue) & 9 & 6.60 (0.29) & 7.87 (0.08) & 8.44 (0.14) \\
N-C$_{10}$H$_{8}^+$ (Exo) (Figure 2-e-blue) & 9 & 6.60 (0.14) & 8.28 (0.28) & 8.40 (0.22) \\
NH-C$_{10}$H$_{8}^+$ (Figure 2-h-blue) & 9 & 6.31 (0.17) & 7.38 (0.11) & 8.42 (0.03) \\
NH$_2$-C$_{10}$H$_{8}^+$ (Figure 2-k-blue) & 9 & 6.28 (0.08) & 7.50 (0.07) & 8.49 (0.03) \\
N-C$_{10}$H$_{8}^+$ (Endo) (Figure 2-n-blue) & 9 & 6.11 (0.10) & 7.31 (0.97) & 8.66 (0.37) \\
H$^+$N-C$_{10}$H$_{8}$ (Exo) (Figure 2-f-blue) & 9 & 6.26 (0.38) & 7.95 (0.35) & 8.68 (0.12) \\
H$^+$NH-C$_{10}$H$_{8}$ (Figure 2-i-blue) & 9 & 6.37 (0.21) & 7.30 (0.17) & 8.71 (0.04) \\
H$^+$NH$_{2}$-C$_{10}$H$_{8}$ (Figure 2-l-blue) & 9 & 6.27 (0.07) & 8.20 (0.33) & 8.47 (0.16) \\
H$^+$N-C$_{10}$H$_{8}$ (Endo) (Figure 2-o-blue) & 9 & 6.44 (0.17) & 7.36 (0.09) & 8.65 (0.05) \\ \hline
Pyrene &  &  & & \\
N-C$_{16}$H$_{10}$ (Exo) (Figure 2-d-red) & 15 & 6.19 (0.02) & 7.90 (0.02) & 8.64 (0.04)  \\
NH-C$_{16}$H$_{10}$ (Figure 2-g-red) & 15 & 6.23 (0.13) & 7.71 (0.04) & 8.61 (0.04) \\
NH$_2$-C$_{16}$H$_{10}$ (Figure 2-j-red) & 15 & 6.13 (0.61) & 7.63 (0.36) & 8.61 (0.22) \\
N-C$_{16}$H$_{10}$ (Endo) (Figure 2-m-red) & 15 & 6.46 (0.62) & 7.93 (0.27) & 8.47 (0.14) \\
N-C$_{16}$H$_{10}^+$ (Exo) (Figure 2-e-red) & 15 & 6.50 (0.43) & 7.40 (0.23) & 8.41 (0.09) \\
NH-C$_{16}$H$_{10}^+$ (Figure 2-h-red) & 15 & 6.23 (0.13) & 7.30 (0.04) & 8.61 (0.04) \\
NH$_2$-C$_{16}$H$_{10}^+$ (Figure 2-k-red) & 15 & 6.26 (0.16) & 7.53 (0.11) & 8.46 (0.07) \\
N-C$_{16}$H$_{10}^+$ (Endo) (Figure 2-n-red) & 15 & 6.14 (0.20) & 7.75 (0.11) & 8.46 (0.05) \\
H$^+$N-C$_{16}$H$_{10}$ (Exo) (Figure 2-f-red) & 15 & 6.29 (0.60) & 7.24 (0.40) & 8.40 (0.16) \\
H$^+$NH-C$_{16}$H$_{10}$ (Figure 2-i-red) & 15 & 6.27 (0.61) & 7.45 (0.30) & 8.43 (0.08) \\
H$^+$NH$_{2}$-C$_{16}$H$_{10}$ (Figure 2-l-red) & 15 & 6.58 (0.27) & 7.44 (0.17) & 8.48 (0.07) \\
H$^+$N-C$_{16}$H$_{10}$ (Endo) (Figure 2-o-red) & 15 & 6.20 (0.15) & 7.81 (0.37) & 8.80 (0.17) \\ \hline
Perylene & & &  & \\
N-C$_{20}$H$_{12}$ (Exo) (Figure 2-d-green) & 19 & 6.33 (0.19) & 7.37 (0.07) & 8.80 (0.02)  \\
NH-C$_{20}$H$_{12}$ (Figure 2-g-green) &19 & 6.22 (0.41) & 7.57 (0.16) & 8.48 (0.07) \\
NH$_2$-C$_{20}$H$_{12}$ (Figure 2-j-green) & 19 & 6.36 (1.12) & 7.84 (1.11) & 8.50 (0.21) \\
N-C$_{20}$H$_{12}$ (Endo) (Figure 2-m-green) & 19 & 6.38 (0.61) & 7.96 (0.43) & 8.80 (0.12) \\
N-C$_{20}$H$_{12}^+$ (Exo) (Figure 2-e-green) & 19 & 6.48 (0.33) & 7.71 (0.37) & 8,44 (0.11) \\
NH-C$_{20}$H$_{12}^+$ (Figure 2-h-green) & 19 & 6.37 (1.12) & 7.78 (0.67) & 8.48 (0.12) \\
NH$_2$-C$_{20}$H$_{12}^+$ (Figure 2-k-green) & 19 & 6.36 (0.08) & 7.83 (1.11) & 8.43 (0.24) \\
N-C$_{20}$H$_{12}^+$ (Endo) (Figure 2-n-green) & 19 & 6.33 (0.41) & 7.59 (0.08) & 8.48 (0.05) \\
H$^+$N-C$_{20}$H$_{12}$ (Exo) (Figure 2-f-green) & 19 & 6.48 (0.47) & 7.89 (0.41) & 8.42 (0.15) \\
H$^+$NH-C$_{20}$H$_{12}$ (Figure 2-i-green) & 19 & 6.37 (1.02) & 7.33 (0.74) & 8.42 (0.20) \\
H$^+$NH$_{2}$-C$_{20}$H$_{12}$ (Figure 2-l-green) & 19 & 6.30 (0.21) & 7.63 (0.31) & 8.47 (0.07) \\
H$^+$N-C$_{20}$H$_{12}$ (Endo) (Figure 2-o-green) & 19 & 6.37 (0.41) & 7.84 (0.38) & 8.42 (0.10) \\\hline
Coronene & &  & & \\
N-C$_{24}$H$_{12}$ (Exo) (Figure 2-d-cyan) & 23 & 6.23 (0.20) & 7.63 (0.19) & 8.89 (0.20) \\
NH-C$_{24}$H$_{12}$ (Figure 2-g-cyan) & 23 & 6.27 (0.20) & 7.63 (0.19) & 8.89 (0.20) \\
NH$_2$-C$_{24}$H$_{12}$ (Figure 2-j-cyan) & 23 & 6.19 (0.47) & 7.78 (0.78) & 8.63 (0.28) \\
 N-C$_{24}$H$_{12}$ (Endo) (Figure 2-m-cyan) & 23 & 6.19 (0.12) & 7.60 (0.37) & 8.50 (0.14) \\
 N-C$_{24}$H$_{12}^+$ (Exo) (Figure 2-e-cyan) & 23 & 6.40 (1.03) & 7.26 (0.70) & 8.43 (0.22) \\
 NH-C$_{24}$H$_{12}^+$ (Figure 2-h-cyan) & 23 & 6.20 (0.59) & 7.42 (0.32) & 8.62 (0.08) \\
 NH$_2$-C$_{24}$H$_{12}^+$ (Figure 2-k-cyan) & 23 & 6.24 (0.51) & 7.41 (0.14) & 8.73 (0.05) \\
 N-C$_{24}$H$_{12}^+$ (Endo) (Figure 2-n-cyan) & 23 & 6.15 (0.72) & 7.52 (0.15) & 8.64 (0.06) \\
 H$^+$N-C$_{24}$H$_{12}$ (Exo) (Figure 2-f-cyan) & 23 & 6.25 (0.81) & 7.35 (0.70) & 8.46 (0.18) \\
 H$^+$NH-C$_{24}$H$_{12}$ (Figure 2-i-cyan) & 23 & 6.23 (0.61) & 7.39 (0.38) & 8.68 (0.13) \\
H$^+$NH$_{2}$-C$_{24}$H$_{12}$ (Figure 2-l-cyan) & 23 & 6.31 (0.35) & 7.65 (0.20) & 8.45 (0.09) \\
	 H$^+$N-C$_{24}$H$_{12}$ (Endo) (Figure 2-o-cyan) & 23 & 6.25 (0.53) & 7.75 (0.28) & 8.42 (0.10) \\\hline
	 Ovalene & & & & \\
		 N-C$_{32}$H$_{14}$ (Exo) (Figure 2-d-yellow) & 31 & 6.21 (0.25) & 7.88 (0.06) & 8.63 (0.04)  \\
		 NH-C$_{32}$H$_{14}$ (Figure 2-g-yellow) & 31 & 6.22 (0.33) & 7.41 (0.21) & 8.62 (0.20) \\
		 NH$_2$-C$_{32}$H$_{14}$ (Figure 2-j-yellow) & 31 & 6.22 (0.76) & 7.76 (0.61) & 8.48 (0.32) \\
		 N-C$_{32}$H$_{14}$ (Endo) (Figure 2-m-yellow) & 31 & 6.21 (0.52) & 7.89 (0.35) & 8.42 (0.33) \\
		 N-C$_{32}$H$_{14}^+$ (Exo) (Figure 2-e-yellow) & 31 & 6.30 (0.79) & 7.50 (0.68) & 8.52 (0.29) \\
		 NH-C$_{32}$H$_{14}^+$ (Figure 2-h-yellow) & 31 & 6.29 (0.53) & 7.38 (0.28) & 8.46 (0.08) \\
		 NH$_2$-C$_{32}$H$_{14}^+$ (Figure 2-k-yellow) & 31 & 6.20 (0.46) & 7.40 (0.17) & 8.50 (0.07) \\
		 N-C$_{32}$H$_{14}^+$ (Endo) (Figure 2-n-yellow) & 31 & 6.13 (0.59) & 7.62 (0.23) & 8.56 (0.07) \\
		 H$^+$N-C$_{32}$H$_{14}$ (Exo) (Figure 2-f-yellow) & 31 & 6.27 (1.03) & 7.46 (0.94) & 8.48 (0.35) \\
		 H$^+$NH-C$_{32}$H$_{14}$ (Figure 2-i-yellow) & 31 & 6.27 (1.10) & 7.43 (0.51) & 8.42 (0.25) \\
		 H$^+$NH$_{2}$-C$_{32}$H$_{14}$ (Figure 2-l-yellow) & 31 & 6.28 (0.42) & 7.30 (0.59) & 8.50 (0.20) \\
		 H$^+$N-C$_{32}$H$_{14}$ (Endo) (Figure 2-o-yellow) & 31 & 6.22 (0.39) & 7.88 (0.22) & 8.62 (0.07) \\\hline
		 C$_{48}$H$_{18}$ & & & & \\
		 N-C$_{48}$H$_{18}$ (Exo) (Figure 2-d-magenta) & 47 & 6.44 (0.33) & 8.00 (0.06) & 8.46 (0.08)  \\
		 NH-C$_{48}$H$_{18}$ (Figure 2-g-magenta) & 47 & 6.41 (0.57) & 7.74 (0.18) & 8.50 (0.14) \\
		 NH$_2$-C$_{48}$H$_{18}$ (Figure 2-j-magenta) & 47 & 6.29 (0.87) & 7.62 (0.54) & 8.49 (0.22) \\
		 N-C$_{48}$H$_{18}$ (Endo) (Figure 2-m-magenta) & 47 & 6.36 (0.77) & 7.46 (0.42) & 8.43 (0.46) \\
		 N-C$_{48}$H$_{18}$ $^+$ (Exo) (Figure 2-e-magenta) & 47 & 6.35 (1.02) & 7.42 (0.80) & 8.52 (0.20) \\
		 NH-C$_{48}$H$_{18}$ $^+$ (Figure 2-h-magenta) & 47 & 6.26 (0.97) & 7.38 (1.02) & 8.59 (0.56) \\
		 NH$_2$-C$_{48}$H$_{18}$ $^+$ (Figure 2-k-magenta) & 47 & 6.43 (0.72) & 7.33 (0.18) & 8.45 (0.11) \\
		 N-C$_{48}$H$_{18}$ $^+$ (Endo) (Figure 2-n-magenta) & 47 & 6.28 (1.44) & 7.46 (0.38) & 8.49 (0.30) \\
		 H$^+$N-C$_{48}$H$_{18}$ (Exo) (Figure 2-f-magenta) & 47 & 6.29 (0.58) & 7.49 (0.93) & 8.43 (0.38) \\
		 H$^+$NH-C$_{48}$H$_{18}$ (Figure 2-i-magenta) & 47 & 6.41 (0.51) & 7.66 (1.92) & 8.49 (0.59) \\
		 H$^+$NH$_{2}$-C$_{48}$H$_{18}$ (Figure 2-l-magenta) & 47 & 6.29 (0.48) & 7.53 (1.00) & 8.45 (0.29) \\
		 H$^+$N-C$_{48}$H$_{18}$ (Endo) (Figure 2-o-magenta) & 47 & 6.44 (0.72) & 7.60 (0.21) & 8.46 (0.14) \\\hline
		 Circumcoronene & &  &  & \\
	$^{\dagger}$C$_{54}$H$_{18}$ (Figure 2-a-black) & . . . & 6.28 (0.22) & 7.68 (0.89) & 8.52 (0.10)  \\
		N-C$_{54}$H$_{18}$ (Exo) (Figure 2-d-black) & 53 & 6.21 (0.24) & 7.70 (0.86) & 8.50 (0.06)  \\
		 NH-C$_{54}$H$_{18}$ (Figure 2-g-black) & 53 & 6.20 (0.27) & 7.71 (0.44) & 8.54 (0.79) \\
		 NH$_2$-C$_{54}$H$_{18}$ (Figure 2-j-black) & 53 & 6.20 (1.02) & 7.64 (1.21) & 8.48 (0.41) \\
		 N-C$_{54}$H$_{18}$ (Endo) (Figure 2-m-black) & 53 & 6.19 (0.43) & 7.79 (1.01) & 8.50 (0.38) \\
	$^{\dagger}$C$_{54}$H$_{18}^+$ (Figure 2-b-black) & . . . & 6.30 (2.37) & 7.85 (0.93) & 8.41 (0.34)  \\
		 N-C$_{54}$H$_{18}^+$ (Exo) (Figure 2-e-black) & 53 & 6.27 (2.29) & 7.87 (1.01) &  8.45 (0.37) \\
		 NH-C$_{54}$H$_{18}^+$ (Figure 2-h-black) & 53 & 6.15 (0.63) & 7.44 (0.98) & 8.46 (0.22) \\
		 NH$_2$-C$_{54}$H$_{18}^+$ (Figure 2-k-black) & 53 & 6.17 (0.88) & 7.41 (1.02) & 8.49 (0.28) \\
		 N-C$_{54}$H$_{18}^+$ (Endo) (Figure 2-n-black) & 53 & 6.14 (1.36) & 7.70 (0.48) & 8.44 (0.19) \\
   $^{\dagger}$H$^+$C$_{54}$H$_{18}$ (Figure 2-c-black) & . . . & 6.22 (1.29) & 7.45 (1.17) & 8.46 (0.59)  \\
		 H$^+$N-C$_{54}$H$_{18}$ (Exo) (Figure 2-f-black) & 53 & 6.20 (1.33) & 7.44 (1.14) & 8.48 (0.37) \\
		 H$^+$NH-C$_{54}$H$_{18}$ (Figure 2-i-black) & 53 & 6.17 (0.52) & 7.47 (0.98) & 8.48 (0.18) \\
		 H$^+$NH$_{2}$-C$_{54}$H$_{18}$ (Figure 2-l-black) & 53 & 6.20 (1.08) & 7.55 (1.19) & 8.42 (0.29) \\
		H$^+$N-C$_{54}$H$_{18}$ (Endo) (Figure 2-o-black) & 53 & 6.15 (1.04) & 7.57 (0.87) & 8.43 (0.31) \\\hline
		\vspace{0.3mm}
		{\raggedright \scriptsize $^{\dagger}$ same data for pure larger PAH}
\end{longtable}
\end{center}

The 6.2 $\mu \rm m$ UIR band has been observed at two wavelengths, either near 6.3 $\mu \rm m$ (longer wavelength component) towards planetary nebulae, post-AGB objects and Herbig AeBe stars or near 6.2 $\mu \rm m$ (shorter wavelength component) towards H~$\textsc{ii}$  regions, reflection nebulae and galaxies (Peeters et al. 2002). The emergence of the shorter wavelength component is an enigma that has been explained by cationized PANHs (Hudgins et al. 2005) and protonated PANHs (Alvaro Galu\'e et al. 2010). This paper shows that the feature at 6.2 $\mu \rm m$ is seen for the neutrals as well. The substitution of N atom into a PAH redistributes its electron density, leading to the changes in the C-C bonds force constants and the dipole derivatives, which further result in a blue-shift for the 6.2 $\mu \rm m$ band. The blueshifting of the 6.2 $\mu \rm m$ band ceases, once the uniformity of the electron distribution over carbon skeletan is attained (Hudgins et al. 2005).

The role of the symmetry among PANHs may not be significant in relation to the peak position of the 6.2 $\mu \rm m$ band, however it seems related to the symmetry of their parent PAH molecule. Table 2 reveals that compact PAHs achieve more blue shifting in the peak position of 6.2 $\mu \rm m$ after inclusion of N atom into the PAH structure. C$_{48}$H$_{18}$, a less symmetrical PAH, shows redder positions compared to more symmetric ones; circumcoronene, ovalene and coronene after N inclusion (Table 2). The shortest wavelength for the 6.2 $\mu \rm m$ band is attained by endo N-PAH ions. Pure circumcoronene cation produces the C-C stretching feature at 6.35 $\mu \rm m$, which shifts to 6.14 $\mu \rm m$ after incorporation of a N atom within the structure (Table 2). Neutral NH$_2$-PAHs produce strong 6.2 $\mu \rm m$ feature that shifts blue-wards with increasing \enquote*{parent} PAH's size and symmetry. For naphthalene, this feature is at 6.43 $\mu \rm m$ and shifts to 6.19 $\mu \rm m$ for circumcoronene (Table 2). Endo N-PAH cations have the band shifting to a shorter wavelength with increasing PAH size. Thus, among all PANH molecules variants, the 6.2 $\mu \rm m$ band shifts towards bluer positions more elegantly with increasing size when N is incorporated within a compact PAH. For ions, the C-C stretching 6.2 $\mu \rm m$ band is stronger in exo N-PAH, NH-PAH and endo N-PAH compared to the remaining variants. The strong C-C stretching feature of protonated endo N-PAHs appears at wavelengths shorter than 6.2 $\mu \rm m$ for circumcoronene (Figure 2 (o)-black).
\begin{table*}
\tbl{Calculated peak positions and intensities ($10^{-13}$W$C^{-1}${$\mu \rm m$}$^{-1}$) of 9.0--15 $\mu \rm m$ features for PAHs and PANH variants in neutral, cationic ($^+$) and protonated forms (H$^+$)}
{\begin{tabular}{lccccc}
\noalign{\vskip3pt}
\hline
PAHs & pure & exo N- & NH- & NH$_2$- & endo N-\\
\hline
	C$_{16}$H$_{10}$ (duo) & 11.89 (0.76) & 12.16 (0.49) & 12.15 (0.52) & 12.22 (0.21) & 11.93 (0.36)\\
	C$_{16}$H$_{10}^+$ (duo) & 11.70 (0.31) & 11.93 (0.21) & 11.93 (0.33) & 11.96 (0.39) & 11.78 (0.58)\\
	H$^+$C$_{16}$H$_{10}$ (duo) & 11.60 (0.10) & 11.98 (0.09) & 11.95 (0.11) & 12.19 (0.26) & 11.98 (0.46)\\\cmidrule{2-6}
	C$_{20}$H$_{12}$ (trio) & 12.50 (0.88)  & 12.42 (0.54) & 13.33 (0.46) & 13.33 (0.19) & 13.16 (0.78)\\
	C$_{20}$H$_{12}^+$ (trio) &  12.51 (0.37) & 12.44 (0.25) & 12.56 (0.45) & 12.96 (0.39) & 12.39 (0.57)\\
    H$^+$C$_{20}$H$_{12}$ (trio) &  12.64 (0.12) & 12.76 (0.13) & 12.42 (0.12) & 12.96 (0.17) & 12.91 (0.20)\\\cmidrule{2-6}
	C$_{24}$H$_{12}$ (duo) & 11.72 (1.44) & 11.78 (0.93) & 11.92 (0.45) & 11.95 (0.20) & 11.92 (0.47)\\
    C$_{24}$H$_{12}^+$ (duo) & 11.61 (0.39) & 11.61 (0.26) & 11.64 (0.40) & 11.69 (0.55) & 11.64 (0.67)\\
	H$^+$C$_{24}$H$_{12}$ (duo) & 11.61 (0.28) & 11.64 (0.19) & 11.69 (0.28) & 11.68 (0.34) & 11.69 (0.57)\\\cmidrule{2-6}
	C$_{32}$H$_{14}$ (solo) & 11.36 (1.23)  & 11.37 (0.72) & 11.39 (0.28) & 11.40 (0.35) & 11.29 (0.35)\\
	C$_{32}$H$_{14}^+$ (solo) & 11.11 (0.35) & 11.10 (0.23) & 11.08 (0.27) & 11.15 (0.48) & 11.19 (0.89)\\
	H$^+$C$_{32}$H$_{14}$ (solo) & 11.23 (0.26) & 11.21 (0.19) & 11.19 (0.21) & 11.09 (0.39) & 11.10 (0.69)\\
	C$_{32}$H$_{14}$ (duo) & 12.03 (0.67) & 12.05 (0.73) & 12.15 (0.12) & 12.20 (0.11) & 12.20 (0.06)\\
	C$_{32}$H$_{14}^+$ (duo) & 11.82 (0.30) & 11.88 (0.29) & 11.95 (0.21) & 11.77 (0.33) & 11.69 (0.65)\\
    H$^+$C$_{32}$H$_{14}$ (duo) & 11.88 (0.17) & 11.81 (0.12) & 11.90 (0.18) & 11.80 (0.11) & 11.77 (0.21)\\\cmidrule{2-6}
	C$_{48}$H$_{18}$ (solo) & 11.04 (2.93)  & 11.10 (2.07) & 11.04 (1.09) & 11.01 (0.62) & 11.08 (1.11)\\
	$C_{48}$H$_{18}^+$ (solo) & 10.98 (0.85) & 10.99 (0.69) & 11.00 (0.75) & 10.91 (1.02) & 10.97 (0.89)\\
    H$^+$C$_{48}$H$_{18}$ (solo) &  10.92 (0.75) & 10.94 (0.59) & 10.98 (0.57) & 10.99 (0.57) & 11.00 (1.94)\\
	C$_{48}$H$_{18}$ (duo) & 11.81 (0.14) & 11.84 (0.27) & 11.67 (0.50) & 11.55 (0.21) & 11.56 (0.19)\\
	C$_{48}$H$_{18}^+$ (duo) & 12.48 (0.24) & 12.57 (0.12) & 11.59 (0.16) & 11.58 (0.25) & 11.87 (0.09)\\
    H$^+$C$_{48}$H$_{18}$ (duo) & 12.48 (0.13) & 12.57 (0.19) & 12.78 (0.26) & 12.22 (0.15) & 12.53 (0.23)\\\cmidrule{2-6}
	C$_{54}$H$_{18}$ (solo) & 10.99 (3.23) & 10.95 (2.18) & 10.95 (1.18) & 11.04 (0.67) & 11.06 (1.19)\\
	C$_{54}$H$_{18}^+$ (solo) & 10.79 (0.78) & 10.80 (0.57) & 10.87 (0.90) & 10.87 (1.08) & 10.76 (1.82)\\
	H$^+$C$_{54}$H$_{18}$ (solo) & 10.76 (0.74) & 10.79 (0.54) & 10.87 (0.62) & 10.87 (0.52) & 10.79 (1.23)\\
	C$_{54}$H$_{18}$ (duo) & 11.81 (0.13) & 11.85 (0.22) & 11.65 (0.52) & 11.58 (0.22) & 11.53 (0.20)\\
	C$_{54}$H$_{18}^+$ (duo) & 12.53 (0.20) & 12.49 (0.14) & 11.65 (0.13) & 11.57 (0.20) & 11.81 (0.07)\\
  H$^+$C$_{54}$H$_{18}$ (duo) & 12.47 (0.18) & 12.41 (0.16) & 13.02 (0.23) & 12.38 (0.16) & 12.42 (0.28)\\
  \hline
  \end{tabular}}
  \begin{tabnote}
  solo, duo and trio stand for the number of C-H groups present in a PAH ring.
  \end{tabnote}
\end{table*}

The peak position of the 7.7 $\mu \rm m$ band does not follow any specific manner with increasing size---this emerges with a blue shifting around 7.5 $\mu \rm m$ for endo N-PAH ions, while it is consistently near 7.6--7.7 $\mu \rm m$ for NH$_2$-PAH neutrals (Table 2). The peak position for the 8.6 $\mu \rm m$ band does not change with the molecular size. PAH cations produce strong 7.7 and 8.6 $\mu \rm m$ features (Peeters et al. 2002; Tielens 2008). In ions, both the bands are strong in the spectra of exo N-PAH and NH-PAH variants. Noticeably, neutral NH$_2$-PAH and neutral endo N-PAH also produce strong 7.7 and 8.6 $\mu \rm m$ bands (Table 2) unlike typical neutral PAHs. The strength of both bands seems to increase with size, and also with the protonation.
\subsection{The C-H out-of-plane Bending Vibrations (9.0--15 $\mu \rm m$)}~
The peak positions and intensities of the C-H out-of-plane bending modes for each selected PANH variant are summarized in Table 1, while their detailed information is listed in Table 3. The corresponding spectra are shown in Figure 2. Figure 1 shows that there are solo, duo and trio C-H groups on an outer ring of the PAHs chosen here. In general, the C-H out-of-plane band for solo C-H groups falls at 10.8--11.4 $\mu \rm m$, while for duo and trio C-H groups, this falls at 11.4--13.2 $\mu \rm m$ (Hudgins \& Allamandola 1999). The peak positions due to C-H out-of-plane bending modes do not shift with the PANH variants. However, these are blue-shifted upon ionization for most PAHs, while red shift is seen due to the C-H out-of-plane band in larger PAHs with duo C-H groups (C$_{48}$H$_{18}$ and C$_{54}$H$_{18}$). The intensities in this region change upon ionization. For neutrals, these are larger in pure, exo N-PAH and NH-PAH variants, but, on the contrary, NH$_2$ and endo N-PAH ions have stronger intensities (Table 1).

The C-H out-of-plane bending feature for PAHs blue-shifts with the increasing number of solo C-H groups that is reported in numerous studies (Bauschlicher et al. 2008, Ricca et al. 2012, Candian \& Sarre 2015), while the physical reason is not well understood. The similar trend is followed by each PANH variant as well. This falls between 11.11--11.40 $\mu \rm m$ for C$_{32}$H$_{14}$ (having 2 solo C-H groups) and 10.76--11.06 $\mu \rm m$ for C$_{54}$H$_{18}$ (having 6 solo C-H groups) (Table 3), whereas their corresponding ions fall at shorter wavelengths than neutrals for every PANH variant considered here. Figure 2 illustrates that spectral features in 9.0-15 $\mu \rm m$ range in PAH (row I-Figure 2), exo N-PAH (row II-Figure 2) and NH-PAH (row III-Figure 2); are stronger for neutrals and weaker for ions (Table 1). However, for NH$_2$-PAH (row IV-Figure 2) and endo N-PAH (row V-Figure 2) neutrals and ions, the spectra tend to produce similar intensities for both C-H out-of-plane bending and C-C stretching vibrations at $\sim$11--14 $\mu \rm m$ and $\sim$6.1--6.4 $\mu \rm m$, respectively (Table 1). The band strengths in the 9.0--15.0 $\mu \rm m$ region increases as a larger number of C-H groups start to contribute in the PANH variant for solo, duo and trio C-H out-of-plane bending vibrations. For instance, the intensity of the C-H out-of-plane band in C$_{54}$H$_{18}$ (6 solo C-H groups) is significantly stronger than C$_{32}$H$_{14}$ (2 solo C-H groups). The intensities due to solo C-H out-of-plane bending vibration are larger than those due to duos.
\begin{figure}
\centering
      \includegraphics[height=10cm, width=8cm]{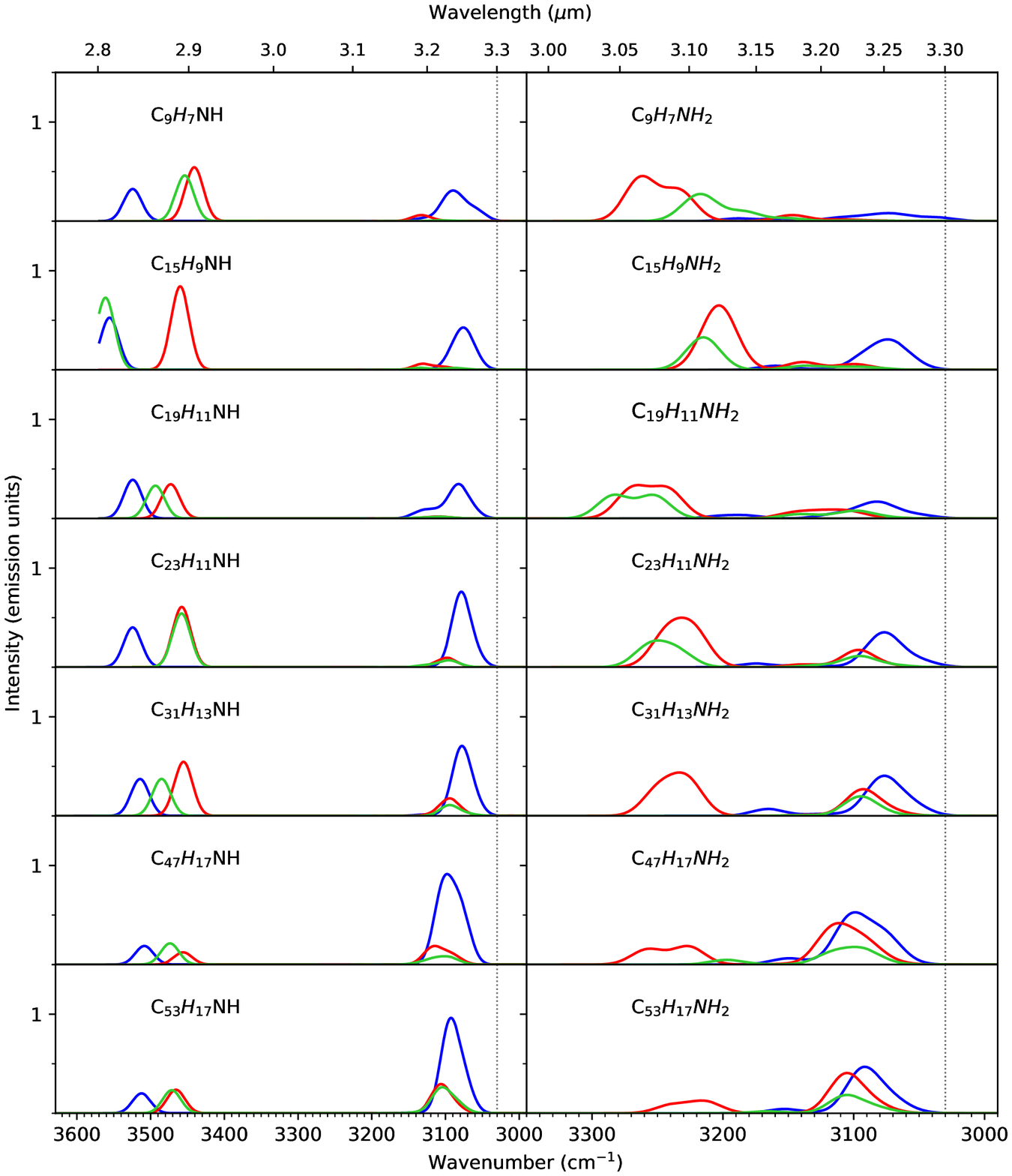}\\
      \caption{Theoretical IR spectra of NH stretching for NH and NH$_2$ bonds in NH-PAH (Column I) and NH$_2$-PAH (Column II) respectively. The blue, red and green show neutral, cationic and protonated forms. The dotted vertical line shows the position of 3.3 $\mu \rm m$ band in celestial objects. The spectra are plotted on an absolute scale with $10^{-13}$W$C^{-1}${$\mu \rm m$}$^{-1}$ intensity unit.}
      \label{Figure 4.}
\end{figure}

\subsection{Vibrations of N-related bonds}~
The C-N stretching and in-plane bending vibrations appear in the 6--9 $\mu \rm m$ range but are extremely weak compared to the regular PAH bands. Therefore, their contribution towards UIR features is suggested to be insignificant.

N-H stretching, on the other hand, gives rise to intense features. Figure 3 shows the spectra for NH-PAH and NH$_2$-PAH, which illustrates the N-H stretching for N-H bond (Column I) and N-H$_2$ bond (Column II) alongside the 3.3 $\mu \rm m$ band. The spectra are given for neutral (blue), cationic (red) and protonated (green) forms. The peak wavelength regions and intensities of N-H stretching modes are summarized in Table 1. N-H stretching vibrations give rise to features near 2.8--2.9 $\mu \rm m$ in NH-PAH, where N is attached with one H atom (Figure 3, Column I). For NH$_2$-PAH, in which N is associated with two H atoms, symmetric and antisymmetric stretching modes of N-H$_2$ bonds produce bands at 3.0--3.1 $\mu \rm m$ (Figure 3, Column II). The N-H stretching features are strong for ions and stronger compared to the C-H stretching features for small PAH ions (C$_{10}$H$_{8}$, C$_{16}$H$_{10}$ and C$_{20}$H$_{12}$). They shift red-wards for NH-PAH ions (Figure 3, Column I). The N-H stretching features are very weak in N-H$_2$ bonds of neutrals and become weaker for large PAHs (Figure 3). The N-H stretch intensity decreases for C$_{48}$H$_{18}$ and C$_{54}$H$_{18}$. We consider here only one N-H bond per PAH and thus the effect of N-inclusion in the distribution of charge becomes less with the increase of the PAH size.
\section{Astrophysical Implications}~
Observations of the UIR features towards different astronomical objects suggest variations in terms of peak position, profile and intensity. These variations in the UIR bands have been grouped into four classes---A, B, C (Peeters et al. 2002; van Diedenhoven et al. 2004) and D (Matsuura et al. 2014). Class A objects show the C-C stretching band at $\sim$6.2 $\mu \rm m$ and the 7.6 $\mu \rm m$ component dominates over the 7.8 $\mu \rm m$ component in the 7.6--7.8 $\mu \rm m$ complex. Class B shows a band around 6.3 $\mu \rm m$ and the 7.8 $\mu \rm m$ is more intense. Classes C and D are characterized by a very broad feature at around 8 $\mu \rm m$, while Class D also shows a broad 6.2 $\mu \rm m$ band. Class A objects are found more prevalent in the observations (Canelo et al. 2018; Pino et al. 2008). Hudgins et al. (2005) conclude the viable existence of N-containing PAHs (PANHs); large size endoskeletal PANHs ($\geq$ 50 C atoms) mainly in Class A objects, while small size exoskeletal PANHs ($\leq$ 50 C atoms) for Class B. Since PANHs considered here do not reproduce the broad features, Classes C and D are not relevant to the present study. 

This work presents the theoretical IR analysis of PAHs ranging in size from naphthalene (8 C atoms) to circumcoronene (54 C atoms) containing nitrogen into their structure for neutrals, cations and protonated forms. Four PANH variants have been studied -- exo N-PAH, NH-PAH, NH$_2$-PAH and endo N-PAH (shown in Figure 1). NH and NH$_2$ embedded in the PAH ring are reported for the first time. Protonated forms of the PANH variants considered here are also studied for the first time as well as the larger size neutrals of exo N-PAHs and endo N-PAHs. The present study extends the previous studies by increasing the sample of PAH species, including larger PAHs, N in the ring structure, which are more relevant to astronomical context.

The mid-IR spectra of exo N-PAH (Fig. 2 and row II) and NH-PAH (Fig. 2 row III) have features similar to pure PAHs, i.e., the C-H stretch features (near 3.3 $\mu \rm m$) and the C-H out of plane features (11 - 13 $\mu \rm m$) are intense for the neutral and the C-C stretch features and C-H in-plane bending features (in the 6-9 $\mu \rm m$ region) are strong for the ions. Apart from the intensity, the position of the features also match well with that of pure PAHs. NH-PAH is an exception as it shows a strong band at 8.54 $\mu \rm m$ for C$_{54}$H$_{18}$. The exo N-PAH variant of C$_{54}$H$_{18}$ matches the Class B positions of the 6.2 $\mu \rm m$ and 7.7 $\mu \rm m$ UIR bands. Whereas, large endoskeletal PANH cations (Hudgins et al. 2005) and protonated PANHs (Alvaro Galu\'e et al. 2010) tend to match the Class A band positions of the UIR bands.

Our results show that the protonated exo N-PAHs and NH-PAHs show a feature at 7.4 $\mu \rm m$, at much shorter wavelength compared to the 7.7 $\mu \rm m$ UIR band. It may be noted that the 7.7 $\mu \rm m$ UIR band also shows a subfeature at 7.4 $\mu \rm m$ (Peeters et al. 2002). We also note that exo N-PAH cations with increasing size and symmetry matches the Class B position of the 6.2 and 7.7 $\mu \rm m$ bands.

In the case of pure PAHs, it is well known that their cations match the intensity of the UIR bands in the 6-9 $\mu \rm m$ region. In the present work, we find that neutral NH$_2$-PAHs and endo N-PAHs show intense features in this region. The intensity increases with the size of these PAHs. In both these variants, the C-H solo out of plane feature (corresponding to the 11.2 $\mu \rm m$ UIR band) is found at longer wavelength for C$_{54}$H$_{18}$ (at $\sim$11.05 $\mu \rm m$) compared to pure PAHs (10.99 $\mu \rm m$). Thus, neutral NH$_2$-PAHs and endo N-PAHs, with more than 50 carbon atoms, match better the Class A position of the 11.2 $\mu \rm m$ band. It should be noted that pure PAHs of similar size have these bands at wavelengths shorter than 11.0 $\mu \rm m$ (Ricca et al. 2012).

The NH$_2$-PAHs and endo N-PAHs (neutrals, cations and protonated forms) also match well the Class A position of the 6.2 $\mu \rm m$ UIR band. A similar result only for endo N-PAH cation is already reported (Hudgins et al. 2005). Apart from cations, our results support the presence of the neutral and protonated forms of these PAHs in the ISM. It is also seen that large neutral NH$_2$-PAHs match better the position of the 7.7 $\mu \rm m$ UIR band, which is not so consistent for the endo-PAHs. While large neutral NH$_2$-PAHs match well the Class A positions of the 6.2, 7.6 and 11.2 $\mu \rm m$ UIR features, endo N-PAHs match the 6.2 and 11.2 $\mu \rm m$ bands fairly well.

The emergence of new features at 2.9 $\mu \rm m$ due to the stretching of N-H bonds is observed for NH-PAH. For ionized NH-PAHs, this feature is as intense as the 3.3 $\mu \rm m$ band. NH$_2$ symmetric and antisymmetric stretching modes in NH$_2$-PAHs produce new features near 3.0--3.1 $\mu \rm m$ for ions. This 3.0 $\mu \rm m$ band is much fainter than the 3.3 $\mu \rm m$ band. Spectra of the Class A \& B objects taken by ISO/SWS (Sloan et al. 2003) and AKARI (Mori et al. 2014) do not show any positive evidence for the presence of features at 2.9--3.0 $\mu \rm m$. If these PAH variants are responsible for the UIR bands, we should be able to detect a band at 2.9 $\mu \rm m$ for NH-PAH, for which we already have a good upper limit. Thus, their contribution must not be very large, while the 3.0 $\mu \rm m$ band of NH$_{2}$-PAH neutrals is very weak and the currently available data do not put a strong constraint or proof for it. IR Spectra of NH and NH$_{2}$ containing PAHs have been reported previously (Bauschlicher 1998) along with CN containing PAHs, concluding that CN and NH$_{2}$ side groups in PAHs do not contribute to the UIR bands. For NH and NH$_{2}$ embedded in the PAH ring as well, the present study confirms the conclusions of Bauschlicher (1998) with a larger sample of NH-PAHs and NH$_{2}$-PAHs.

On the other hand, exo N-PAHs (with N at the periphery without H) and endoskeletal PANHs do not show features in 2.9--3.1 $\mu \rm m$ since they do not have N-H bonds on the periphery and there is no constraint on their contribution. 
While N-PAHs with N at the periphery match with the peak positions of Class B UIR bands, it has affinity to bond with H or H$^{+}$, suggesting that the N incorporation inside the structure might be more relevant. The absence of N-H stretch features in spectra of celestial objects can be accounted for either by N present in exo N-PAHs (with N at the periphery without H) or endo N-PAHs (with N inside the structure) or low nitrogen abundance in PAHs. The presence of PANHs should not be ignored as the recent discovery of CN-containing PAH (McGuire et al. 2021) marks a silver-lining for their presence in the ISM.
\begin{figure}
\centering
      \includegraphics[height=10cm,width=8cm]{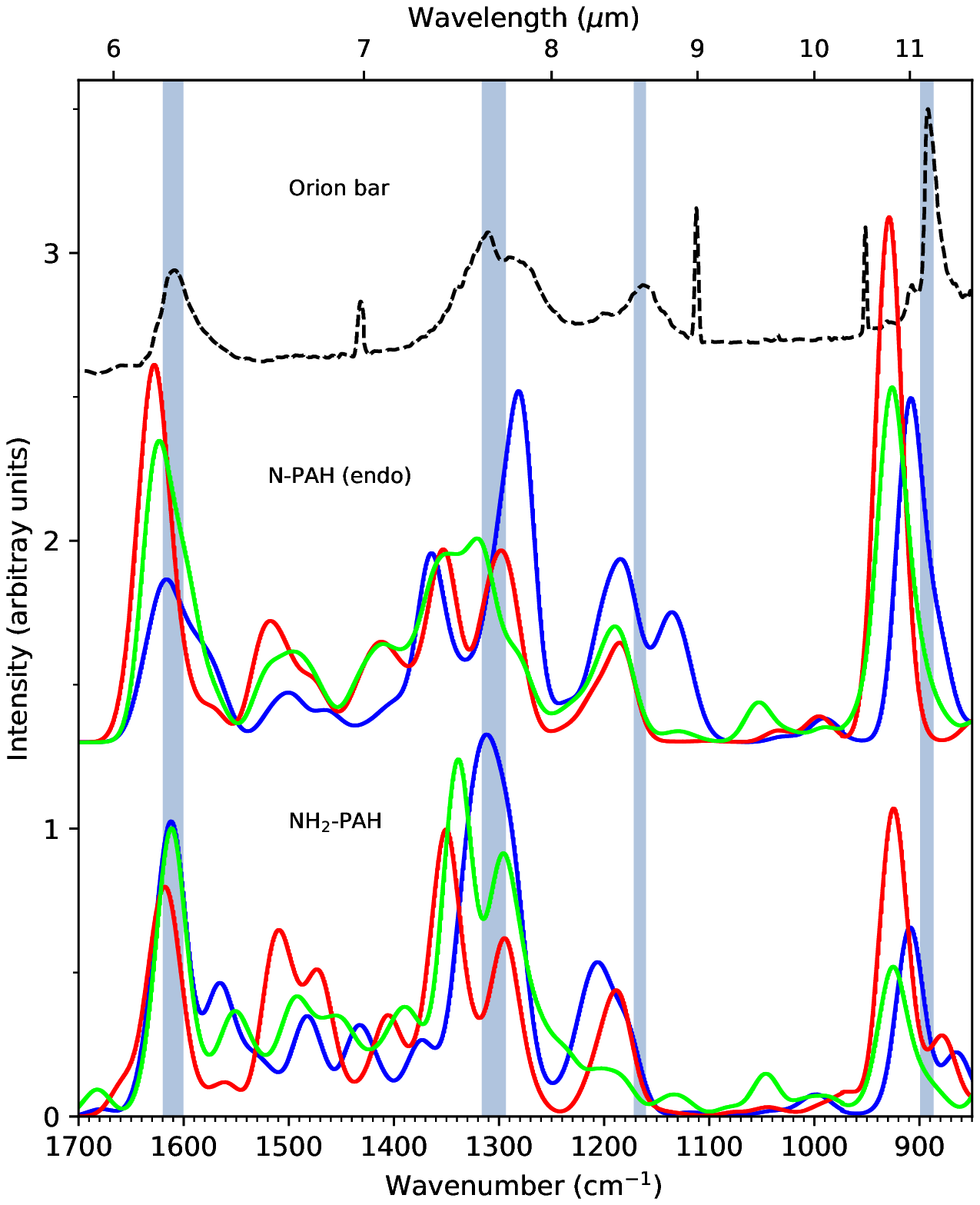}\\
      \caption{Theoretical mid-IR spectra of circumcoronene in neutral (blue), cation (red) and protonation (green). The dashed line represents the spectra of the Orion bar taken from Sloan et al. (2003). The shaded vertical region spans the range of the observed position of the 6.2, 7.7 and 8.6 $\mu \rm m$ bands, as Peeters et al. (2002) and the 11.2 $\mu \rm m$ band, as van Diedenhoven et al. (2004). }
      \label{Figure 4.}
\end{figure}

Observations reveal that bands in the 6--9 $\mu \rm m$ region and the 11.2 $\mu \rm m$ arise from different populations (Galliano et al. 2008). The ascription of the 3.3 and 11.2 $\mu \rm m$ bands with neutral PAHs and those in the 6-9 $\mu \rm m$ region with ionized PAHs is well established (Peeters et al. 2002; Tielens 2008; Schmidt et al. 2009; Rigopoulou et al. 2021). Despite this, various correlations have been found between PAH bands along with some distinct spatial morphologies (Peeters et al. 2017; Sidhu et al. 2021; Knight et al. 2021), which intimates more complexity in the variations between PAH bands. The ionization fraction is determined by the balance between photoionization and recombination and this may account for the observed variations in relative band intensities. These variations also suggest that factors other than ionization may play a role in the appearance of the UIR bands.

Figure 4 compares the mid-IR spectra of Orion bar with endoskeletal PANHs and with NH$_{2}$-PAHs (with N at the periphery). We clearly see a significant resemblance of the theoretically calculated spectra with the observed one. The Orion bar spectra has almost equally strong 6.2 and 11.2 $\mu \rm m$ bands with a very strong 7.7 $\mu \rm m$ band. Several of the PAHs shown in Figure 4 exhibit similar spectral behaviour.

The present study shows that neutral endoskeletal PANHs have the strong 6.2 and 11.2 $\mu \rm m$  band together. Thus, if they contribute significantly to the observed UIR bands, then there should be some N--rich astronomical regions where the 6.2 to 11.2 $\mu \rm m$ band ratio may not always be a direct indicator of the ionization degree of PAHs as has been assumed in previous studies.
The spatial variations of such regions could involve various effects, including ionization and nitrogen-inclusion. Interstellar environments, where the 6.2 and 11.2 $\mu \rm m$ bands are equally strong, endo N-PAH may be prospective candidates PAHs to inhabit such environments.
\section{Summary}~
We have calculated four N-containing variants of seven PAHs (ranging in size from 10 carbon atoms up to 54) with N, NH and NH$_{2}$ incorporation in neutral, cationic and protonated forms. Large exo N-PAH and endo N-PAH cations have been considered previously (Hudgins et al. 2005). Small PAHs with NH and NH$_{2}$ as side group have also been studied (Bauschlicher 1998), concluding their negligible contribution to the UIR bands. In the present study, NH and NH$_{2}$ have been incorporated in the outer ring of the PAH molecules.

Neutral and protonated PANHs also match well some of the UIR bands along with PANH cations (Hudgins et al. 2005), suggesting that the interstellar PANH population could be a mixture of these. The present work covers all the general UIR bands and their possible attribution with PANHs, while Hudgins et al. (2005) focus mainly on the 6.2 $\mu \rm m$ band. The present work shows that PANH may contribute significantly to the 11.2 $\mu \rm m$ band in some interstellar regions. PANHs may also contribute to the 7.7 and 8.6 $\mu \rm m$ complex but a more detailed study is required to analyse the quantitative contribution. The evident conclusions of this study are summarized: 

\vspace{1mm}
1. Ionization shows higher intensities for the bands in the 6--9 $\mu \rm m$ region for exo N-PAH and NH-PAH variants and intensities become stronger with increasing size, which is consistent here with previous findings (Hudgins et al. 2005). The present study confirms them for a larger sample of PAH species. The present study also suggests that exo N-PAH cations may show the Class B UIR positions of the 6.2 and 7.7 $\mu \rm m$ bands. 

2. The spectra of PANH neutrals in the 6--9 $\mu \rm m$ region show larger intensities, when N is bonded with two H atoms (NH$_2$-PAH) and also when N is located within the ring (endo N-PAH). This behavior for endo N-PAHs is reported for the first time, and for NH$_2$-PAHs, this agrees with Bauschlicher (1998).  The present study confirms their conclusions for larger size of PAHs and also for NH$_2$ being in the ring, suggesting further that NH$_2$-PAH and endo N-PAHs with larger size (C atoms $>$ 50) may produce 6.2, 7.7 and 11.2 $\mu \rm m$ features as Class A UIR bands.

3. New features at $\sim$2.9 $\mu \rm m$ and 3.0--3.1 $\mu \rm m$ arise due to N-H stretching in NH-PAH and NH$_2$-PAH respectively. Both features look stronger than the 3.3 $\mu \rm m$ band for NH-PAH and NH$_2$-PAH ions (Figure 3), while the 3.0-3.1 $\mu \rm m$ band of neutral NH$_2$-PAH is very faint and sensitive observations are required to detect it. They have not been detected in observations as yet, although sensitive observations available in this range may be limited.

4. The absence of N-H stretching features in observations confines the contribution of NH-PAHs and NH$_2$-PAHs that is consistent with the conclusions of Bauschlicher (1998) for larger size PAHs as well.

5. Large endo N-PAH neutrals and ions produce strong 6.2 and 11.2 $\mu \rm m$ bands. This behavior is also present in the spectra of NH$_2$-PAHs, suggesting that there might be some regions of nitrogen dominance in the ISM where the 6.2 and 11.2 $\mu \rm m$ band ratio loses its exclusive role to be an indicator for the ionizations of PAHs.

6. Endo N-PAH cations with increasing size account for the Class A position of the 6.2 $\mu \rm m$ band, which confirms the conclusion of Hudgins et al. (2005) for cations. Here, we found that protonated (additional H$^{+}$ at C) endo N-PAHs with increasing size and symmetry of its parent PAH molecule also show the same behavior as cations.

\vspace{1mm}

N, being at the periphery of PAHs has an affinity to bond with H or H$^{+}$ to form NH-PAH and NH$_{2}$-PAH. The absence of their corresponding stretching features in observations constrains the contribution of NH and NH$_{2}$ related PAHs. On the other hand, endoskeletal PANHs are free from this constraint. Further theoretical and experimental investigations are required for extensive understanding.
\section*{Acknowledgements}
We are thankful to the anonymous reviewer for comments that helped in bringing clarity to the manuscript. AV acknowledges research fellowship from DST SERB EMR grant, 2017 (SERB-EMR/2016/005266). AP acknowledges financial support from DST SERB EMR grant, 2017 (SERB-EMR/2016/005266), IoE incentive grant, BHU (incentive/2021-22/32439), Banaras Hindu University, Varanasi and thanks the Inter-University Centre for Astronomy and Astrophysics, Pune for associateship. The authors also acknowledge support from DST JSPS grant (DST/INT/JSPS/P-238/2017). MB acknowledges JSPS for research support (Fellowship ID P19029). TO is supported by a Grant-in-Aid for Scientific 
Research of JSPS (18K03691).

\end{document}